\documentclass[twocolumn]{aastex63}

\received{***}
\revised{***}
\accepted{***}
\submitjournal{ApJ}

\shorttitle{Age estimation of supernova remnants}
\shortauthors{Suzuki et al.}

\usepackage{threeparttable}
\usepackage{multirow}
\usepackage{color}
\usepackage{amsmath}
\usepackage{amsfonts}

\begin{document}

\title{Quantitative Age Estimation of Supernova Remnants and Associated Pulsars}

\correspondingauthor{Hiromasa Suzuki}
\email{hiromasa050701@gmail.com}

\author{Hiromasa Suzuki}
\affiliation{Department of Physics, The University of Tokyo,
7-3-1 Hongo, Bunkyo-ku, Tokyo 113-0033, Japan}

\author{Aya Bamba}
\affiliation{Department of Physics, The University of Tokyo,
7-3-1 Hongo, Bunkyo-ku, Tokyo 113-0033, Japan}
\affiliation{Research Center for the Early Universe, The University of Tokyo, 
7-3-1 Hongo, Bunkyo-ku, Tokyo 113-0033, Japan}

\author{Shinpei Shibata}
\affiliation{Department of Physics, Yamagata University, Kojirakawa 1-4-12, Yamagata, 990-8560, Japan}



\begin{abstract}
The age of a supernova remnant (SNR) is, though undoubtedly one of the most important properties for study of its evolution, difficult to estimate reliably in most cases.
In this study, we compare the dynamical and plasma ages of the SNRs and characteristic ages of their associated pulsars with the corresponding SNRs' ages that are generally thought to be reliable ($t_{\rm r}$): historical and light-echo ages of the SNRs, { kinematic ages of the ejecta knots} and kinematic ages of the associated neutron stars (NS).
The kinematic age of { ejecta knots} or a NS is the time that they have taken to reach the current positions from the explosion center.
We use all of the available { 24} systems for which $t_{\rm r}$ is already available (historical, light-echo, and { ejecta kinematic ages}) or measurable (NS kinematic age).
We estimate the NS kinematic ages for eight SNR-NS systems by determining quantitatively the geometric centers of the SNR shells.
The obtained $t_{\rm r}$ ranges from 33~yr to $\approx 400$~kyr.
We find that the two SNR ages, dynamical and plasma ages, are consistent with $t_{\rm r}$ within a factor of four, whereas the characteristic ages of the pulsars differ from $t_{\rm r}$ by more than a factor of four in some systems.
Using the $t_{\rm r}$ summarized in this work, we present the initial spin periods of the associated pulsars, which are more strictly constrained than the previous works, as well.
\end{abstract}



\section{Introduction}

The age of a supernova remnant (SNR) is undoubtedly one of the most important properties for studies of its { thermal and non-thermal evolution}.
However, estimation of the SNR age involves a considerable amount of uncertainty in most cases.
Currently, seven methods to estimate the ages of SNRs are known and the estimated ages in these methods are called historical and light-echo ages, kinematic age of the ejecta knots, dynamical and thermal-plasma ages, characteristic age of the associated pulsar, and kinematic age of the associated neutron star (NS).

The most reliable age estimation among these is based on historical records, i.e., the historical age.
Several celestial transient events that are suspected to be supernova explosions were recorded with their sky positions and time in the human history, and the remnants of some of them have been successfully identified, such as the SNR for SN1006, described in {\it Meigetsuki} written by Fujiwara no Teika (1162--1241) \citep{iba34, reizei03}.
Nine SNRs are recognized as historical SNRs \citep{green03}.

Another reliable age estimation is based on light echoes (light-echo age).
Bright optical emission of a supernova is scattered by interstellar dust in the vicinity and is sometimes observed with a delay of tens to hundreds of years.
The light-echo ages have been determined for six SNRs so far \citep{krause05, rest05, krause08, rest08a, rest08b}.

{ If proper motions of the multiple ejecta knots are measured, a reliable age, the kinematic age of the ejecta knots ($t_{\rm kin, ej}$) can be calculated based on their proper motions and the separation angles between the knots and location of the supernova explosion (the explosion point).
The explosion point can be calculated if the proper motions of multiple knots are measured and thus their converging point is derived.
Currently, $t_{\rm kin, ej}$ is available for six young to middle-aged SNRs \citep{thorstensen01, fesen06, sato17, banovetz21, law20, winkler09, winkler88}}

For most of the SNRs including ones without reliable age estimations, their ages can be derived using the shock dynamics, called dynamical ages, $t_{\rm dyn}$.
The simplest estimation is based on the shock velocity $v_{\rm sh}$ and radius, providing that $v_{\rm sh}$ is known \citep{sedov59}.
But even when $v_{\rm sh}$ is unknown, $t_{\rm dyn}$ can be estimated from  the ambient density $n_{\rm e}$ determined with X-ray observations combined with the SNR radius, which has been accurately determined for many SNRs \citep{sedov59, truelove99}.
In some previous studies, $v_{\rm sh}$ was derived from the post-shock temperature instead of $n_{\rm e}$ or directly measured $v_{\rm sh}$ (e.g., \citealt{vink10}).
It should be noted, however, that $t_{\rm dyn}$ is usually calculated assuming simple density distributions (uniform or power-law function of the radius) and thus the calculated value has potentially significant uncertainty.
In summary, these three ways are possible to calculate $t_{\rm dyn}$. Which method is suitable depends on the available physical properties, i.e., the shock velocity, density, and post-shock temperature, in addition to the radius.
{ The $t_{\rm dyn}$ includes a considerable (and often cannot be quantified) amount of uncertainty associated with the distances, ambient density profiles, and/or plasma volumes.}

The other age estimation which is available for many SNRs is the plasma age ($t_{\rm p}$).
In general, the plasma of an SNR is in a non-equilibrium ionization (NEI) state since it has been gradually ionized after the explosion \citep{itoh89, zhang19}, and they are called ionizing plasmas.
Thus, in principle, once the ionization timescale $n_{\rm e}t$ of the plasma, which can be measured from X-ray spectroscopy, 
has been determined, the plasma age $t_{\rm p}$ is easily calculated, using the separately measured plasma density $n_{\rm e}$.
We should note that this estimation method is in practice applicable only to relatively young SNRs with the age of less than $\sim 10^4$~yr because the time to reach collisional ionization equilibrium is $\sim 10^4$~yr for the typical interstellar medium (ISM) density of 1~cm$^{-3}$.
We also note that the ambient density of the actual systems are not uniform and therefore the current average density may not be appropriate to determine the true age from.

The other population of the NEI plasmas in SNRs besides ionizing plasmas is recombining plasmas, which is the plasma in a recombination-dominant state.
Although its origin is unclear, recent studies suggest that they are also produced in an early stage of the SNR evolution \citep{itoh89, suzuki18, katsuragawa19, zhang19, suzuki20b}.
In this work, the recombination timescale is treated as $n_{\rm e} t_{\rm p}$ as well.
{ The $t_{\rm p}$ suffers from a considerable (and often cannot be quantified) amount of uncertainty associated with the distances, plasma volumes, and ambient density profiles.}

If an SNR is associated with a fast-rotating NS (i.e., pulsar) that was created in the supernova and if the spin period $P$ and period derivative $\dot{P}$ of the pulsar are measured, the pulsar's age $t$, which is by definition the same as the SNR age, is generally given by the equation \citep{manchester77}
\begin{equation}\label{eq-initialspin}
t =\left\{\begin{aligned}
&\frac{2 \tau_{c}}{n-1}\left[1-\left(\frac{P(0)}{P}\right)^{n-1}\right] \quad &\text { for } n \neq 1 \\
&2 \tau_{c} \ln \left(\frac{P(0)}{P}\right) \quad &\text { for } n=1,
\end{aligned}\right.
\end{equation}
where $\tau_{\rm c} = P/2\dot{P}$ is the characteristic age of the pulsar and $P(0)$ and $n$ are the initial spin period and braking index, respectively.
The parameter $\tau_{\rm c}$ is equal to $t$ if $P(0) \ll P$ and $n = 3$.
In general, the actual age may differ from $\tau_{\rm c}$ depending on the values of $P(0)$ and $n$.
As an example, the Crab nebula is the remnant of SN~1054 (0.966~kyr), but $\tau_{\rm c}$ of the Crab pulsar is $\approx 1.24$~kyr (e.g., \citealt{popov12}).
Usually, since $P(0)$ and $n$ are unknown, $\tau_{\rm c}$ is used as a possible indicator of the SNR age.

If the proper motion of the NS associated with an SNR is known, the kinematic age of the system $t_{\rm kin, NS}$ can be derived from the proper motion and separation angle between the NS and location of the explosion point.
Since the ISM density is too low to affect the motion of a NS, the NS should have been traveling with a constant velocity since its birth. Then, the only remaining uncertainty is the explosion point, which is not necessarily trivial to determine. 
The simplest guess is to regard the geometric center of an SNR as the explosion point.
Some previous works estimated $t_{\rm kin, NS}$ using geometric centers determined by eye (CTB~80 by \citealt{migliazzo02, zeiger08}; S~147 by \citealt{kramer03}).
These estimations can be applied in a meaningful way only for the systems where the separation between the presumable geometric center and associated NS is very large.

In general, although some amount of the offset of the explosion point from the geometric center is expected according to hydrodynamic simulations to reproduce observed morphologies of some SNRs (e.g., \citealt{orlando07, lu20}), the amount seems to be reasonably small ($\lesssim 24\%$ of the SNR radius).
We note that it is currently unknown whether an offset as large as 24\% of the radius is present in actual SNR systems because the two required parameters of ambient density-gradient and degree of asymmetry of the supernova explosion have never been determined quantitatively with observations.

\cite{katsuda18} performed the first quantitative estimation of the geometric centers of six SNRs.
Their method (hereafter, K18 method) was in a word to calculate the center positions of a shell along the south-north and east-west directions.
The K18 method is only valid for SNRs with complete shells observed.

In summary, the reliability of none of the two kinds of the estimated SNR ages $t_{\rm dyn}$ and $t_{\rm p}$ and the associated pulsar's characteristic age $\tau_{\rm c}$, which is a potential indicator of the SNR age, have been evaluated quantitatively.
In this work, we investigate their accuracy in a quantitative manner for the first time, using the SNR systems with measurable reliable ages $t_{\rm r}$, which is one (or more) of the historical age, light-echo age, { $t_{\rm kin, ej}$}, and $t_{\rm kin, NS}$.
First, we measure $t_{\rm kin, NS}$ for our sample by estimating the geometric centers of the SNR shells with an improved method from the K18 method (Section~\ref{sec-kin}).
 Then, for each object of the sample, we compare $t_{\rm dyn}$, $t_{\rm p}$, and $\tau_{\rm c}$ with $t_{\rm r}$ to evaluate their accuracy quantitatively (Section~\ref{sec-comparison}).
Finally, in Section~\ref{sec-period},  we calculate the initial spin periods of the associated pulsars $P(0)$ in the SNRs with associated pulsars in our sample using $t_{\rm r}$, which is an application of the accurately determined ages of SNR-pulsar systems.
{ Throughout the paper, errors in the figures and tables correspond to a 1~$\sigma$ confidence range}.

\section{Sample Selection}\label{sec-sample}
 The sample of this work includes all SNRs with known historical or light-echo ages, { $t_{\rm kin, ej}$}, or measurable $t_{\rm kin, NS}$.
The plasma age $t_{\rm p}$ is also taken from literature; in the cases where only spatially-resolved results are presented in the literature, the average of the X-ray spectroscopy results is regarded as $t_{\rm p}$.
{ We basically adopt the objects' parameters such as the ages and distances presented in the latest works with fewer assumptions in their measurements or calculations (e.g., for the $t_{\rm dyn}$ of N~103B, we adopt \cite{williams18} instead of \cite{ambrocio97} because the proper motion study is generally less uncertain).}
Table~\ref{tab-ages-simple} lists our sample with the flags to indicate which parameters are presented in literature.
The SNRs with measurable $t_{\rm kin, NS}$ in the table are those of which the proper motion of the associated NS and the shapes are known.
We will calculate $t_{\rm kin, NS}$ of these systems by estimating the explosion center coordinates in Section~\ref{sec-kin}.
The selection of these systems are based on \cite{popov12, ng07, hobbs05}.
We will then compare $t_{\rm dyn}$, $t_{\rm p}$, and $\tau_{\rm c}$ with $t_{\rm r}$ in Section~\ref{sec-comparison}.
In addition to the sources listed in Table~\ref{tab-ages-simple}, we use PKS~1209$-$51, RCW~103, and Kes~73 for only verification of our method of explosion-point estimation in Section~\ref{sec-kin}. We consider that these three sources are not suitable for estimation of $t_{\rm kin, NS}$ performed in Section~\ref{sec-kin} with the following reasons. 
In the case of PKS~1209$-$51, the associated pulsar (1E~1207.4$-$5209) has been found not to be moving in the opposite direction to the expected geometric center \citep{halpern15}.
In the case of RCW~103 and Kes~73, although the proper motions of their respective associated pulsars 1E~161348$-$5055 and 1E~1841$-$045 were measured by \cite{holland17},
 the reported proper motions were too well constrained in spite of very small displacements of the NSs, as pointed out by \cite{katsuda18}. 
In addition, the directions of their proper motions contradict the expectations.

\begin{table*}
\fontsize{5}{8}\selectfont
\centering
\caption{
Sample used in this work.
\label{tab-ages-simple}}
\begin{threeparttable}
\begin{tabular}{l l l l l l l l l l l l}
\hline\hline
SNR & Neutron star & Historical & Light echo & { $t_{\rm kin, ej}$} & $\tau_{\rm c}$ & $t_{\rm dyn}$ & $t_{\rm p}$ & $t_{\rm kin, NS}$\tnote{a} & $t_{\rm kin, NS, prev}$\tnote{b} & { $P_2/P_0$\tnote{c}} & $P_3/P_0$\tnote{d} \\
\hline
SN1987A & --- & $ \checkmark$ & $ \checkmark$ & --- & --- & $ \checkmark$ & $ \checkmark$ & --- & --- & --- & --- \\
Cassiopeia~A & --- & $ \checkmark$ & $ \checkmark$ & $\checkmark$ & --- & $ \checkmark$ & $ \checkmark$ & $ \checkmark$ & --- & $2.74 \pm 0.04$ & $5.47 \pm 0.06$ \\
Kepler & --- & $ \checkmark$ & --- & $\checkmark$ & --- & $ \checkmark$ & $ \checkmark$ & --- & --- & $2.28 \pm 0.11$ & $2.10 \pm 0.18$ \\
Tycho & --- & $ \checkmark$ & $ \checkmark$ & --- & --- & $ \checkmark$ & $ \checkmark$ & --- & --- & $0.09 \pm 0.01$ & $0.65 \pm 0.01$ \\
3C58 & J0205+6449 & $ \checkmark$ & --- & --- & $ \checkmark$ & $ \checkmark$ & --- & --- & --- & --- & --- \\
Crab & B0531+21 & $ \checkmark$ & --- & --- & $ \checkmark$ & --- & --- & --- & --- & --- & ---\\
SN1006 & --- & $ \checkmark$ & --- & --- & --- & $ \checkmark$ & $ \checkmark$ & --- & --- & --- & ---\\
G11.2$-$0.3 & J1813$-$1749 & $ \checkmark$ & --- & --- & $ \checkmark$ & $ \checkmark$ & --- & --- & --- & $8.30 \pm 0.32$ & $6.37 \pm 0.46$ \\
RCW86 & --- & $ \checkmark$ & --- & --- & --- & $ \checkmark$ & $ \checkmark$ & --- & --- & $4379 \pm 2$ & $2098 \pm 7$ \\
0509$-$67.5 & --- & --- & $ \checkmark$ & --- & --- & $ \checkmark$ & $ \checkmark$ & --- & --- & $1.78 \pm 1.77$ & $4.08 \pm 3.18$ \\
0519$-$69.0 & --- & --- & $ \checkmark$ & --- & --- & $ \checkmark$ & $ \checkmark$ & --- & --- & $2.19 \pm 1.13$ & $0.81 \pm 0.65$ \\
N~103B & --- & --- & $ \checkmark$ & --- & --- & $ \checkmark$ & $ \checkmark$ & --- & --- & $1.52 \pm 0.34$ & $1.60 \pm 0.65$ \\
1E~0102.2$-$7219 & --- & --- & --- & $\checkmark$ & --- & $ \checkmark$ & $ \checkmark$ & --- & --- & --- & --- \\
N~132D & --- & --- & --- & $\checkmark$ & --- & $ \checkmark$ & $ \checkmark$ & --- & --- & $0.64 \pm 0.25$ & $63.7 \pm 4.2$ \\
G292.0+1.8 & J1124$-$5916 & --- & --- & $\checkmark$ & $\checkmark$ & $ \checkmark$ & $ \checkmark$ & --- & --- & $28.5 \pm 1.5$ & $10.4 \pm 1.4$ \\
Puppis~A & J0821$-$4300 & --- & --- & $\checkmark$ & $ \checkmark$ & $ \checkmark$ & $ \checkmark$ & $ \checkmark$ & --- & --- & --- \\
IC443 & CXOUJ061705.3+222127 & --- & --- & --- & --- & $ \checkmark$ & $ \checkmark$ & This work & --- & --- & --- \\
W44 & B1853+01 & --- & --- & --- & $ \checkmark$ & $ \checkmark$ & $ \checkmark$ & This work & --- & $575 \pm 2$ & $27.0 \pm 0.2$ \\
Vela & B0833$-$45 & --- & --- & --- & $ \checkmark$ & $ \checkmark$ & $ \checkmark$ & This work & --- & --- & --- \\
G114.3+0.3 & B2334+61 & --- & --- & --- & $ \checkmark$ & $ \checkmark$ & --- & This work & --- & $128 \pm 1$ & $1.66 \pm 0.07$ \\
CTB80 & B1951+32 & --- & --- & --- & $ \checkmark$ & $ \checkmark$ & --- & This work & $ \checkmark$ & $20060 \pm 186$ & $2073 \pm 24$ \\
W41 & B1830$-$08 & --- & --- & --- & $ \checkmark$ & $ \checkmark$ & --- & This work & --- & $951 \pm 5$ & $94.5 \pm 7$ \\
S147 & J0538+2817 & --- & --- & --- & $ \checkmark$ & $ \checkmark$ & --- & This work & $ \checkmark$ & --- & --- \\
Monogem & B0656+14 & --- & --- & --- & $ \checkmark$ & $ \checkmark$ & --- & This work & --- & --- & --- \\
\hline
\end{tabular}
\begin{tablenotes}
\item[a] Checkmarks indicate the measurable $t_{\rm kin, NS}$ from the explosion center coordinates which has been accurately obtained in literature based on ejecta kinematics.
\item[b] Checkmarks indicate that $t_{\rm kin, NS}$ from the explosion center coordinates inferred by eye has been reported in literature.
\item[c] { Quadrupole power ratio of the emission structure (radio continuum or infrared), which reflects the mirror asymmetry.}
\item[d] { Octupole power ratio of the emission structure (radio continuum or infrared), which reflects the ellipticity.}
\item[]{\bf Notes.} The available values are indicated by checkmarks. The values are obtained in Section~\ref{sec-kin} for the objects indicated with ``This work''. See Table~\ref{tab-ages} for the references for individual age estimations. PKS~1209$-$51, RCW~103, and Kes~73, which are used in the explosion-center estimation (Section~\ref{sec-kin}), are excluded from the list, for they are not used for any other purposes.
{ The quadrupole and octupole power ratio values are taken from the radio continuum study, \cite{stafford19} (Tycho, RCW~86, W~44, G114.3+0.3, CTB~80, and W~41) and the infrared study, \cite{peters13} (Cassiopeia~A, Kepler, G11.2$-$0.3, 0509$-$67.5, 0519$-$69.0, N~103B, N~132D, and G292.0+1.8).}
\end{tablenotes}
\end{threeparttable}
\end{table*}
\normalsize

\section{Estimation of the Kinematic Ages of the associated Neutron Stars}\label{sec-kin}
 For the calculation of $t_{\rm kin, NS}$, the separation angle between the explosion point of an SNR and its associated NS is required to be determined.
 In this section, we determine the geometric centers of SNRs, which are likely, and thus are assumed in the later sections in this paper, to correspond to the explosion centers.

\subsection{Method}
In the ideal situation, where a symmetric supernova explosion occurs in a uniform ambient density, the SNR will have a circular shock structure and
 the geometric center can be easily determined.
In reality, however, the observed SNR shapes are not simply circular. Thus, in this study, we model-fit the projected morphology of the SNR shell with an ellipse  as a simple enough yet better alternative to a circle
 and estimate the geometric center.
This method works for SNRs with parts of the shell missing as well, in contrast to the K18 method.

The observed shape of each SNR is determined as follows.
 In principle, a radio continuum image is used if available. If no clear radio continuum images are available for the source, an infrared image (specifically, for\ CTB~80) or X-ray images (for Monogem and Vela) are adopted.
Then, a contour level which best represents the SNR shape is selected
 by eye because it is difficult to define a generalized quantitative procedure in order to identify the boundary between the SNR shell and ISM.
 We note that parts of the shell are kept missing in some SNRs if determining the boundaries is difficult due to the low flux.

\begin{figure}[htb]
\centering
\includegraphics[width=8cm]{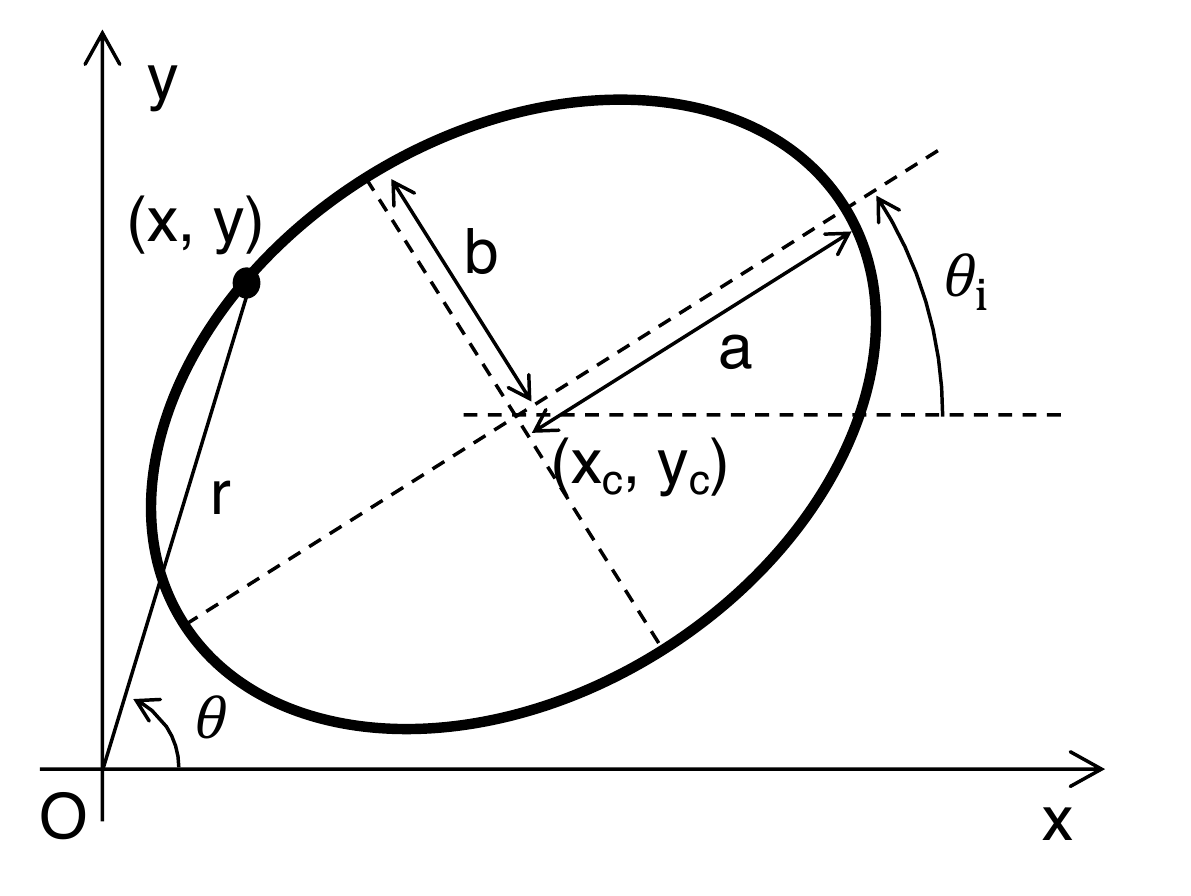}
\caption{Definition of the parameters of the ellipse used in Eqs.~2--3.
\label{fig-ellipse}}
\end{figure}

 We fit the extracted SNR shape with an ellipse as follows.
The extracted contour data consist of nearly evenly-spaced points on the celestial coordinates (R.A., DEC).
 A major difficulty in the fitting is that it is impossible to determine a priori which parts of the complex structures of the contour are suitable and which are not for the fitting. Hence, we reduce  the data points into evenly-spaced $[N / n]$ points, where $N$ is the total point number and $n$ is the reduction level, and fit each of them.
In each reduction level, $n$ ways of data selection are performed.
In this study, we adopt a range of $n$ between $n = 1$ and $n_{\rm max} = [N/10]$, which yields a number of the total data sets of
\begin{equation}
\sum_{n=1}^{n_{\rm max}} n = \frac{1}{2} n_{\rm max} (n_{\rm max} + 1).
\end{equation}
Then, the ellipse for the fitting function on the $x$-$y$ plane is given by
\begin{equation}
\left[\frac{x - (x_{\rm c} \cos\theta_{\rm i} + y_{\rm c}\sin\theta_{\rm i})}{a}\right]^2 + \left[\frac{y - (-x_{\rm c}\sin\theta_{\rm i} + y_{\rm c}\cos\theta_{\rm i})}{b}\right]^2 = 1,
\end{equation}
 where ($x_{\rm c}$, $y_{\rm c}$) are the center of the ellipse, $a$ and $b$ are the major- and minor-axes radii, and $\theta_{\rm i}$ is the angle between the major axis of the ellipse and x-axis. 
 In practice, we transform it into the polar coordinates ($r$, $\theta$) in our fitting, as given by
\begin{equation}
\begin{aligned}
&r (\theta) = \sqrt{ \frac{1}{A} \left(1 - C + \frac{B^2}{A}\right)} + \frac{B}{A},\\
&A = \frac{\cos^2\theta}{a^2} + \frac{\sin^2\theta}{b^2},\\
&B = \frac{(x_{\rm c} \cos\theta_{\rm i} + y_{\rm c}\sin\theta_{\rm i}) \cos\theta}{a^2} + \frac{(-x_{\rm c}\sin\theta_{\rm i} + y_{\rm c}\cos\theta_{\rm i}) \sin\theta}{b^2},\\
&C = \frac{(x_{\rm c} \cos\theta_{\rm i} + y_{\rm c}\sin\theta_{\rm i})^2}{a^2} + \frac{(-x_{\rm c}\sin\theta_{\rm i} + y_{\rm c}\cos\theta_{\rm i})^2}{b^2}.
\end{aligned}
\end{equation}
Fig.~\ref{fig-ellipse} illustrates the parameter definition.
The fitting on each object yields $\frac{1}{2} n_{\rm max} (n_{\rm max} + 1)$ sets of solutions in total.
We adopt the mean and standard deviation of the distribution as the best-fit geometric center coordinates and their uncertainty, respectively.

\subsection{Results}
\begin{figure*}[htb]
\centering
\includegraphics[width=16cm]{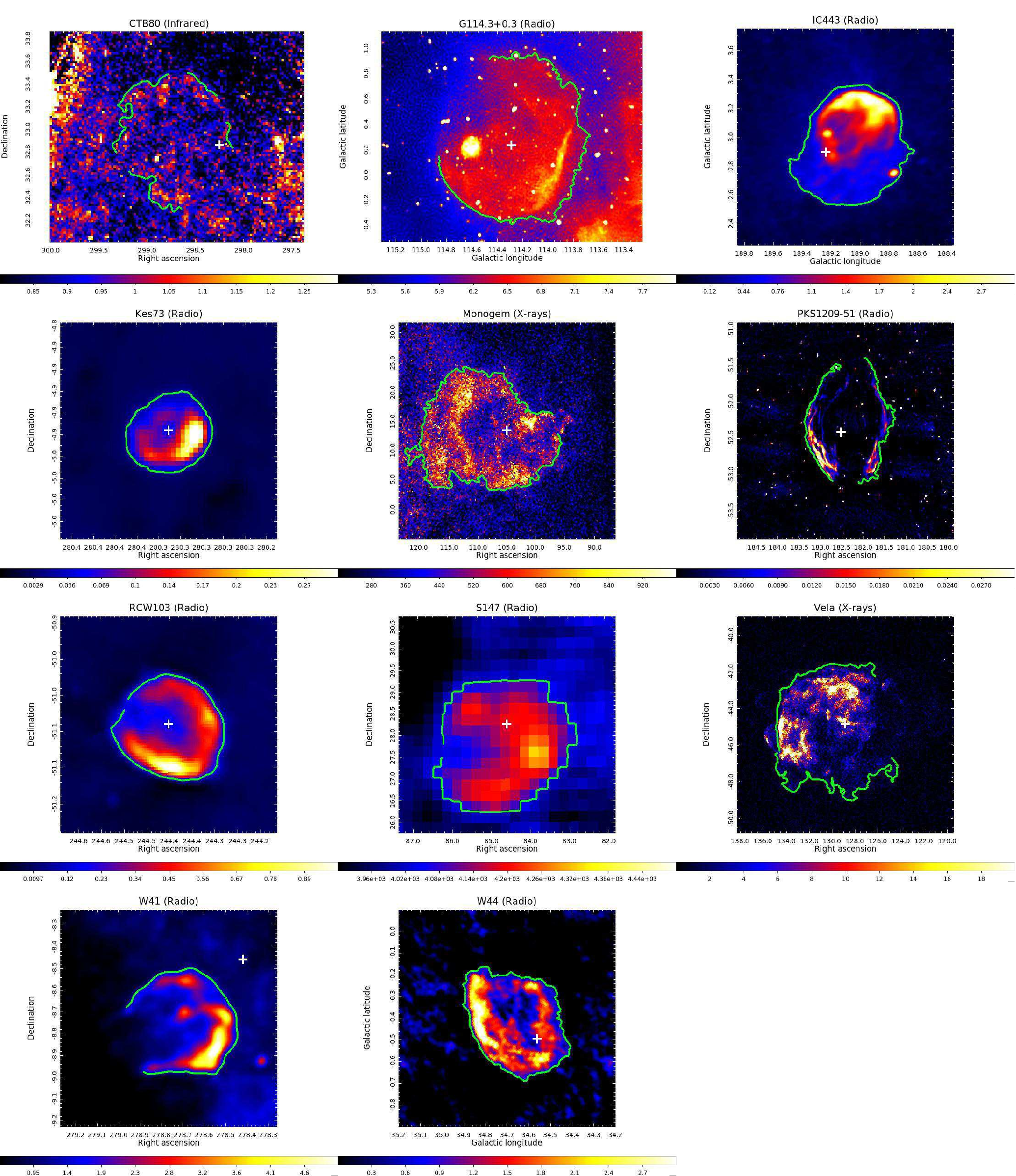}
\caption{Radio, infrared, or X-ray images of the sample in this study. The parts of the contour that is assumed to show the shape of the SNR shell are shown with green lines in each panel.
The white crosses indicate positions of the associated NSs.
\label{fig-snr-psr-img}}
\end{figure*}

\begin{figure*}[htb]
\centering
\includegraphics[width=20cm, angle=90]{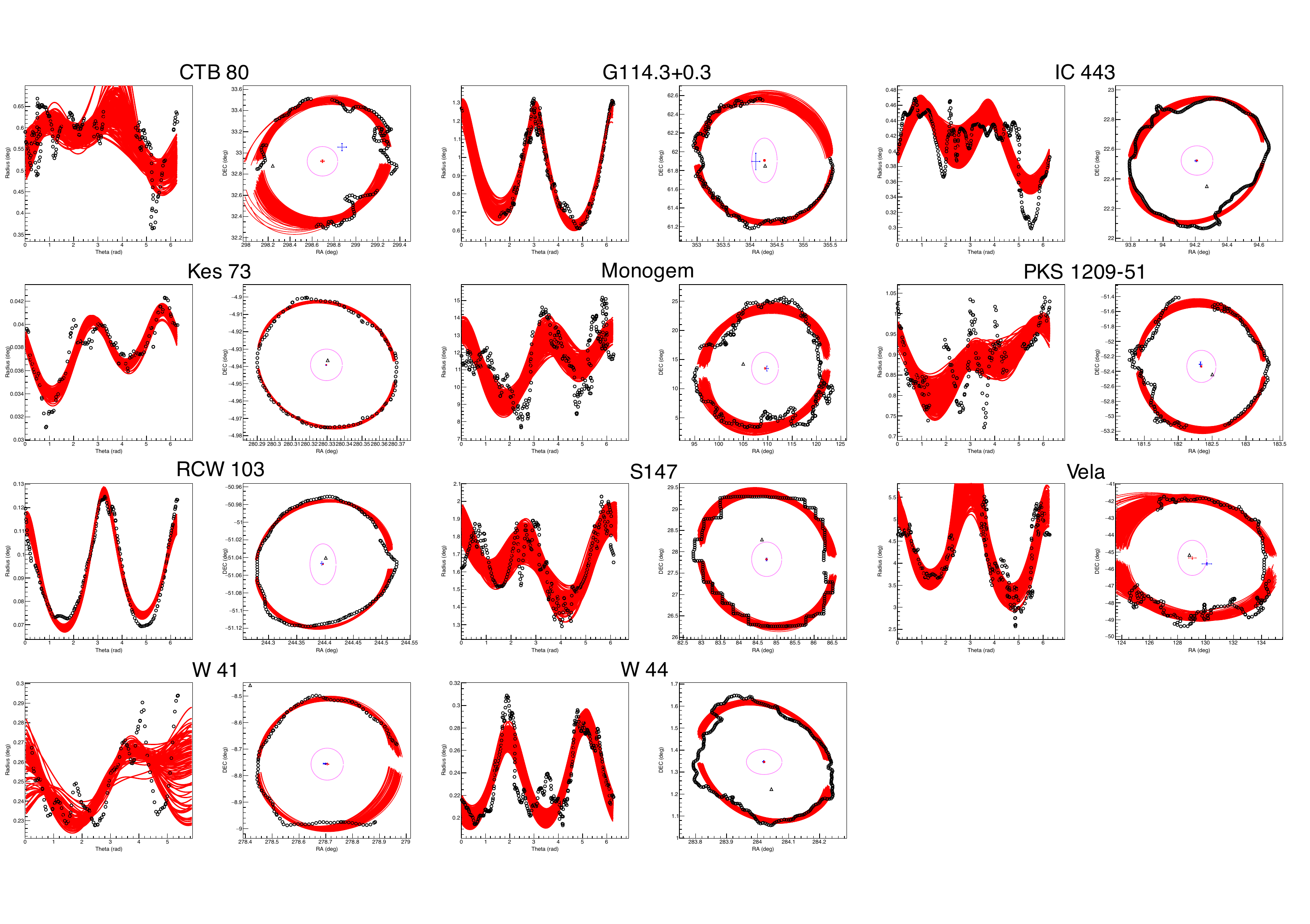}
\caption{For each object, the left panel shows the SNR shape data (black) and the ellipse models (red) on the ($r$, $\theta$) plane and the right panel shows those on the (R.A., DEC) plane.
Red and blue crosses are the best-fit geometric centers and those obtained with the K18 method.
Black triangles give the NS positions at present.
Pink small circle in each panel is centered at the red cross and has a radius of the 24\% average radius of the best-fit shape data, which represents the possible offset between the determined geometric center and actual explosion point \citep{orlando07, lu20}.
\label{fig-snr-psr-fit}}
\end{figure*}

Fig.~\ref{fig-snr-psr-img} shows the images and extracted contours which are used for the geometric center estimation, whereas
Fig.~\ref{fig-snr-psr-fit} shows the ellipse-fitting results.
   We note that the actual explosion point in a SNR may be offset from the determined geometric center by up to 24\% of its radius \citep{orlando07, lu20}, which is also indicated in Fig.~\ref{fig-snr-psr-fit}.
Tables~\ref{tab-coord} and \ref{tab-ages} tabulate the fitting results of the 
 coordinates of the SNR geometric centers and of the NSs and the other parameters, including $t_{\rm kin, NS}$, respectively. Table~\ref{tab-ages} also summarizes selected key parameters of the sample SNRs. 
In consequence, we find that whereas the geometric centers derived with our method are in very good agreement with those with the K18 method only for the complete-shell SNRs, i.e., IC~443, Kes~73, and W~44, 
 they differ greatly  for the SNRs with parts of the shell missing (e.g., CTB~80), as expected.

It should be noted that, in many cases (CTB~80, IC~443, Monogem, S~147, W~41, and W~44), the angular distance between the estimated geometric center of the SNR and associated NS is larger than the possible offset between the geometric center and actual explosion point.
 Therefore, in these cases, $t_{\rm kin, NS}$ estimated here should be reliable even if the latter offsets are taken into account.

\begin{table*}[htb]
\fontsize{7}{10}\selectfont
\centering
\caption{
 Obtained coordinates of the SNR geometric center and associated NS.
\label{tab-coord}}
\begin{threeparttable}
\begin{tabular}{l l r r r r r}
\hline\hline
SNR & pulsar & \multicolumn{2}{l}{NS coordinate ($^ \circ$)\tnote{a}} & \multicolumn{2}{l}{Geometric center coordinate ($^ \circ$)} \\ 
 & & R.A.{ (2000.)} & DEC{ (2000.)} & R.A.{ (2000.)} & DEC{ (2000.)} & { $\Delta\theta$ (arcmin)\tnote{b}} \\
\hline
IC443 & CXOUJ061705.3+222127 & $94.2719$ & $22.3506$ & $94.2109 \pm $0.0042  & $22.5239 \pm $0.0032 & $10.93 \pm $0.20 \\
W44 & B1853+01 & $284.0444$ & $1.2226$ & $284.0210 \pm $0.0022  & $1.3465 \pm $0.0026 & $7.56 \pm $0.15 \\
Vela & B0833-45 & $128.8359$ & $-45.1764$ & $129.0460 \pm $0.2656  & $-45.3451 \pm $0.0721 & $13.46 \pm $8.08 \\
G114.3+0.3 & B2334+61 & $354.2740$ & $61.8504$ & $354.2610 \pm $0.0147  & $61.9081 \pm $0.0084 & $3.48 \pm $0.50 \\
CTB80 & B1951+32 & $298.2425$ & $32.8779$ & $298.6950 \pm $0.0174  & $32.9211 \pm $0.0126 & $22.94 \pm $0.88 \\
W41 & B1830-08 & $278.4178$ & $-8.4588$ & $278.7050 \pm $0.0057  & $-8.7567 \pm $0.0037 & $24.69 \pm $0.28 \\
S147 & J0538+2817 & $84.6044$ & $28.2859$ & $84.9903 \pm $0.0268  & $27.6937 \pm $0.0204 & $28.47 \pm $1.29 \\
Monogem & B0656+14 & $104.9508$ & $14.2387$ & $109.4200 \pm $0.1792  & $13.5087 \pm $0.1746 & $263.98 \pm $41.10 \\
RCW103 & 1E161348-5055 & $244.4013$ & $-51.0403$ & $244.3960 \pm $0.0008  & $-51.0475 \pm $0.0003 & $0.48 \pm $0.09 \\
PKS1209-51 & 1E1207.4-5209 & $182.5037$ & $-52.4412$ & $182.3420 \pm $0.0077  & $-52.3338 \pm $0.0101 & $8.75 \pm $0.49 \\
Kes73 & 1E1841-045 & $280.3304$ & $-4.9365$ & $280.3300 \pm $0.0002  & $-4.9391 \pm $0.0002 & $0.16 \pm $0.01 \\
\hline
\end{tabular}

\begin{tablenotes}
\item[a] Taken from the ATNF pulsar catalog (\url{http://www.atnf.csiro.au/research/pulsar/psrcat/}) or SIMBAD database (\citealt{wenger00}; \url{http://simbad.u-strasbg.fr/simbad/})
\item[b] Separation angle between the geometric center and position of the associated neutron star.
\end{tablenotes}
\end{threeparttable}
\end{table*}

\begin{table*}[htb]
\fontsize{6}{9}\selectfont
\centering
\movetabledown=2.7 in
\begin{rotatetable}
\caption{
Properties of the sample SNRs and their associated NSs.
\label{tab-ages}}
\begin{threeparttable}
\begin{tabular}{l l l l l l l l l l l l l}
\hline\hline
SNR & Neutron star & Historical & Light echo & $t_{\rm kin, ej}$ & $\tau_{\rm c}$ & $t_{\rm dyn}$ (kyr) & $t_{\rm p}$ (kyr) & Velocity\tnote{b} & $\Delta\theta$ (arcmin)\tnote{c} & Distance & $t_{\rm kin, NS}$ (kyr) & $t_{\rm kin, NS, prev}$ \\
 & & age (kyr) & age (kyr) & (kyr) & (kyr)\tnote{a} & & & (km s$^{-1}$) & & (kpc) & & (kyr)\tnote{d} \\
\hline
SN1987A & --- & 0.033 & 0.0339$\pm$0.0014 & --- & --- & 0.059$\pm$0.001 & 0.035$\pm$0.003 & --- & --- & 48 & --- & ---  \\
Cassiopeia~A & --- & 0.340 & --- & 0.336$\pm$0.015 & — & 0.48$\pm$0.04 & 1.1 & 332.0$\pm$20.0 & 0.11$\pm$0.01 & 3.4 & 0.32$\pm$0.03\tnote{e} & ---  \\
Kepler & --- & 0.416 & --- & 0.475$\pm$0.074 & --- & 0.40$\pm$0.06 & 0.09$\pm$0.05 & --- & --- & 4.2 & --- & ---  \\
Tycho & --- & 0.448 & --- & --- & --- & 0.58$\pm$0.05 & $< 0.073$ & --- & --- & 3 & --- & ---  \\
3C58 & J0205+6449 & 0.839 & --- & --- & 5.4 & 2.5 & --- & 35.0$\pm$6.0 & --- & 3.2 & --- & ---  \\
Crab & B0531+21 & 0.966 & --- & --- & 1.24 & --- & --- & 140.0$\pm$8.0 & --- & 2.2 & --- & ---  \\
SN1006 & --- & 1.014 & --- & --- & --- & 1.23$\pm$0.13 & 1.23$\pm$0.06 & --- & --- & 2.2 & --- & ---  \\
G11.2-0.3 & J1813$-$1749 & 1.634 & --- & --- & 23.2 & 0.96$\pm$0.20 & --- & $< 108$ & --- & 4.4 & --- & ---  \\
RCW~86 & --- & 1.835 & --- & --- & --- & 1.84$\pm$0.59 & 1.10$\pm$0.76 & --- & --- & 2.5 & --- & ---  \\
0509$-$67.5 & --- & --- & 0.4$\pm$0.15 & --- & --- & 0.31$\pm$0.04 & 0.4 & --- & --- & 48 & --- & ---  \\
0519$-$69.0 & --- & --- & 0.6$\pm$0.2 & --- & --- & 0.45$\pm$0.20 & 0.3 & --- & --- & 48 & --- & ---  \\
N~103B & --- & --- & 0.86$\pm$0.48 & --- & --- & 0.85 & $> 1.2$ & --- & --- & 48 & --- & ---  \\
1E~0102.2$-$7219 & --- & --- & --- & 1.738$\pm$0.175 & --- & 2.10$\pm$0.50 & 1.60$\pm$0.40 & --- & --- & 60.6 & --- & ---  \\
N~132D & --- & --- & --- & 2.450$\pm$0.195 & --- & 5.75$\pm$1.45 & 9.31$\pm$1.90 & --- & --- & 48 & --- & ---  \\
G292.0+1.8 & J1124$-$5916 & --- & --- & 2.990$\pm$0.060 & 2.85 & 2.60$\pm$0.25 & 2.85$\pm$1.60 & --- & --- & 6 & --- & ---  \\
Puppis~A & J0821$-$4300 & --- & --- & 3.7$\pm$0.3 & 1489 & 4 & 7.42$\pm$0.48 & 763.0$\pm$73.0 & 6.11$\pm$0.49 & 2 & 4.58$\pm$0.57\tnote{e} & ---  \\
IC~443 & CXOUJ061705.3+222127 & — & --- & --- & --- & 4 & 15.77$\pm$1.31 & $< 372$ & 10.93$\pm$0.20 & 1.5 & $> 12.62$ & ---  \\
W~44 & B1853+01 & --- & --- & --- & 20.4 & 55.00$\pm$20.00 & 16.70$\pm$2.50 & 348.5$\pm$69.2 & 7.56$\pm$0.15 & 3 & 18.63$\pm$3.73 & ---  \\
Vela & B0833$-$45 & --- & --- & --- & 11.3 & 9.50$\pm$2.50 & 3.47$\pm$0.19 & 59.5$\pm$2.0 & 13.46$\pm$8.08 & 0.294 & 19.02$\pm$11.46 & ---  \\
G114.3+0.3 & B2334+61 & --- & --- & --- & 40.6 & 7.7 & --- & $< 35.9$ & 3.48$\pm$0.50 & 0.7 & $> 19.42$ & ---  \\
CTB~80 & B1951+32 & --- & --- & --- & 107 & 60 & --- & 273.0$\pm$11.0 & 22.94$\pm$0.88 & 2 & 48.09$\pm$2.68 & 58$\pm$7  \\
W~41 & B1830$-$08 & --- & --- & --- & 148 & 100 & --- & 543.7$\pm$96.3 & 24.69$\pm$0.28 & 4.6 & 59.77$\pm$10.64 & ---  \\
S~147 & J0538+2817 & --- & --- & --- & 618.1 & 140.00$\pm$60.00 & --- & 105.2$\pm$30.0 & 28.47$\pm$1.24 & 1.47 & 113.81$\pm$32.92 & 37$\pm$4  \\
Monogem & B0656+14 & --- & --- & --- & 110.9 & 128.00$\pm$42.00 & --- & 60.0$\pm$7.0 & 263.98$\pm$10.46 & 0.28 & 353.18$\pm$43.65 & ---  \\
\hline
\end{tabular}

\begin{tablenotes}
\item[a] Characteristic age of the pulsar taken from \cite{popov12} except for W41, for which the value is taken from \cite{hobbs05}.
\item[b] Velocity of the NS taken from \cite{ng07} or \cite{hobbs05} except for Puppis~A and IC443, for which the values are taken from \cite{mayer20} and \cite{swartz15}, respectively.
\item[c] Separation angle between the geometric center and position of the associated neutron star.
\item[d] Parameter $t_{\rm kin, NS}$ reported in literature (\citealt{migliazzo02, zeiger08} for CTB80; \citealt{kramer03} for S147).
\item[e] Calculated from the coordinates of the explosion derived by \citealt{thorstensen01, fesen06} (for Cassiopeia~A) and \citealt{winkler88} (for Puppis~A).
\item[]{\bf Notes.} The references for ($t_{\rm dyn}$; $t_{\rm p}$; $t_{\rm kin, ej}$; Distance) of individual sources are:
SN1987A:(\citealt{frank16, park06}; --- ; \citealt{panagia99});
Cassiopeia~A:(\citealt{patnaude09, murray79} ; \citealt{thorstensen01}, \citealt{fesen06}; \citealt{reed95});
Kepler:(\citealt{katsuda08a}; \citealt{katsuda15a}; \citealt{sato17} ; \citealt{katsuda08a});
Tycho:(\citealt{hughes00, hwang02}; --- ; \citealt{tian11}, \citealt{hayato10});
3C58:(\citealt{chevalier05}, \citealt{bietenholz06}, \citealt{rudie07}; --- ; --- ; \citealt{roberts93});
Crab:( --- ; --- ; --- ; \citealt{trimble68}, \citealt{trimble73});
SN1006:(\citep{winkler14, yamaguchi08} ; --- ; \citealt{winkler03});
G11.2-0.3:(\citealt{tam03}, \citealt{borkowski16}; --- ; --- ; \citealt{green04});
RCW86:(\citealt{helder13, lemoine12}; --- ; \citealt{rosado96}, \citealt{sollerman03});
0509$-$67.5:(\citealt{hovey15}; \citealt{kosenko08}; --- ; \citealt{macri06});
0509$-$69.0:(\citealt{kosenko10}; \citealt{kosenko10}; --- ; \citealt{macri06});
N~103B:(\citealt{williams18}; \citealt{kosenko08}; --- ; \citealt{macri06});
1E~0102.2$-$7219:(\citealt{xi19}; \citealt{xi19}; \citealt{banovetz21}; \citealt{hilditch05});
N~132D:(\citealt{hughes87}; \citealt{bamba18}; \citealt{law20}; \citealt{macri06});
G292.0+1.8:(\citealt{gonzalez03}; \citealt{gonzalez03}, \citealt{kamitsukasa14}; \citealt{winkler09}; \citealt{gaensler03});
Puppis~A:(\citealt{blair95, petre82}; --- ; \citealt{reynoso03});
IC~443:(\citealt{troja08, matsumura18}; --- ; \citealt{ambrocio17}, \citealt{zhao20});
W~44:(\citealt{suzuki20b, uchida12}; --- ; \citealt{lee19}, \citealt{wang20});
Vela:(\citealt{aschenbach95}, \citealt{sushch11}; --- ; \citealt{bocchino99, dodson03});
G114.3+0.3:(\citealt{yar-uyaniker04}; --- ; --- ; \citealt{yar-uyaniker04}, \citealt{zhao20});
S~147:(\citealt{sofue80}, \citealt{kundu80}; --- ; --- ; \citealt{ng07b});
W~41:(\citealt{tian07}; --- ; --- ; \citealt{lee19}, \citealt{wang20});
CTB~80:(\citealt{koo90}, \citealt{leahy12}; --- ; --- ; \citealt{strom00});
Monogem:(\citealt{plucinsky96}, \citealt{thorsett03}; --- ; --- ; \citealt{plucinsky96})

\end{tablenotes}
\end{threeparttable}
\end{rotatetable}
\end{table*}
\normalsize

\section{Comparison of the Ages Estimated with Individual Methods}\label{sec-comparison}
The ages estimated with a variety of methods for each SNR, i.e., historical age, light-echo age, { $t_{\rm kin, ej}$}, $t_{\rm dyn}$, $t_{\rm p}$, $\tau_{\rm c}$, and $t_{\rm kin, NS}$, have been summarized in Table~\ref{tab-ages}.
Ideally, they all would agree with each other. In this section, we verify whether they agree and if so, look into to which extent.

\begin{figure*}[htb]
\centering
\includegraphics[width=16cm, angle=0]{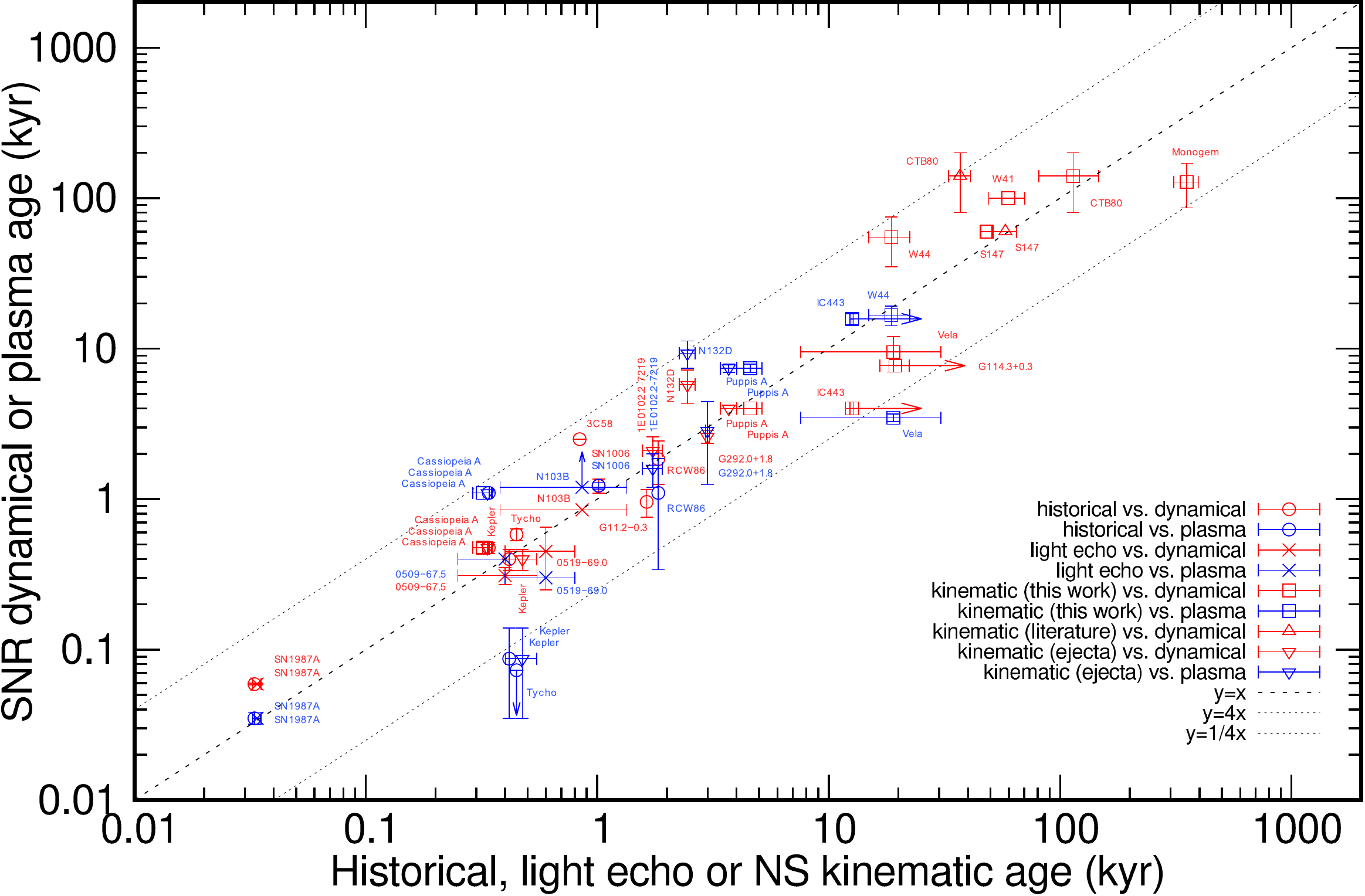}
\caption{Comparison of a variety of age estimation results.
The $x$-axis represents $t_{\rm r}$ (circles for historical ages, crosses for light-echo ages, inverted triangles for $t_{\rm kin, ej}$, squares for $t_{\rm kin, NS}$, triangles for $t_{\rm kin, NS, prev}$).
The $y$-axis represents the dynamical (red) or plasma (blue) ages of SNRs.
For reference, the linear functions $y=x, 4x$, and $1/4 x$ are also plotted.
\label{fig-ages1}}
\end{figure*}

\begin{figure*}[htb]
\centering
\includegraphics[width=16cm, angle=0]{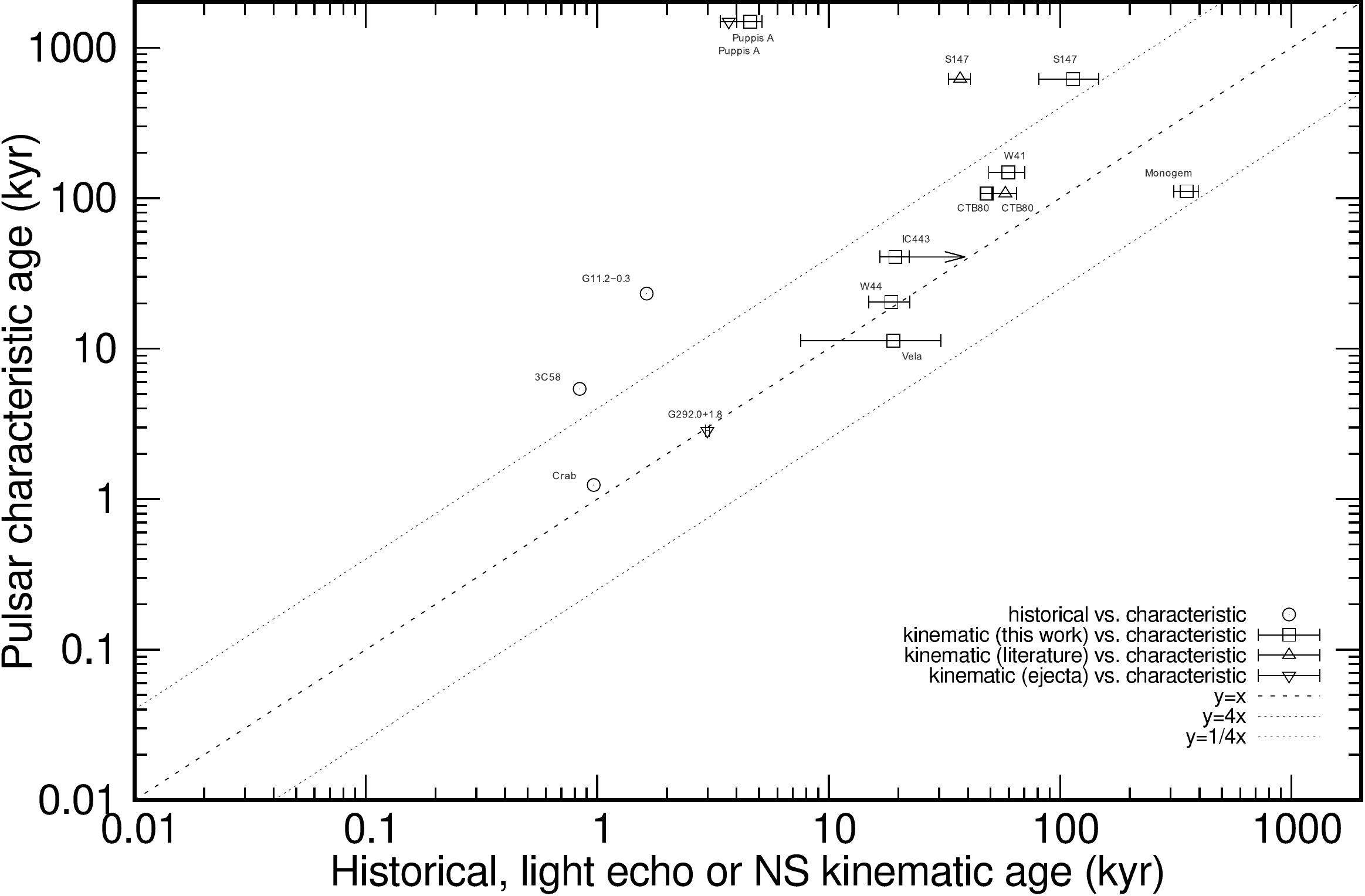}
\caption{ Same as Fig.~\ref{fig-ages1} except that the $y$-axis represents $\tau_{\rm c}$.
\label{fig-ages2}}
\end{figure*}

Fig.~\ref{fig-ages1} plots $t_{\rm dyn}$ and $t_{\rm p}$ as a function of $t_{\rm r}$ (historical ages, light echo ages, and/or $t_{\rm kin, NS}$) and Fig.~\ref{fig-ages2} does $\tau_{\rm c}$.
Ideally, all the points would be on the straight line on the $x$-$y$ plane ($y = x$). 

As for $t_{\rm dyn}$ and $t_{\rm p}$, the values of the most SNRs in our sample are in good agreement with their $t_{\rm r}$ within a factor of four.
The errors of $t_{\rm dyn}$ and $t_{\rm p}$ cannot be generally quantified. However, we conjecture that they might be attributed simply to the errors of the distance $d$ and volume filling factor of the SNR plasma $f$ according to the relation $t_{\rm dyn} \propto d^{2.5} f^{-0.25}$ \citep{sedov59} and $t_{\rm p} \propto d^{-1.5} f^{0.5}$ (e.g., \citealt{itoh89}).
If that is the case, the obtained dispersion of a factor of four at most corresponds to the errors of factors of $< 1.8$ and $< 20$ for $d$ and $f$, respectively.
This constraint for $d$ seems to be reasonable because the typical errors of $d$ have been argued to be $\lesssim 10\%$ (e.g., \citealt{zhao20}), whereas the observational dispersion of $f$ is completely unknown.
As for $\tau_{\rm c}$, although the values of eight SNRs in our sample agree well with their $t_{\rm r}$, five objects show very large discrepancies (a factor of 5--330) from their $t_{\rm r}$.
This tendency is more or less expected because $\tau_{\rm c}$ is larger than the actual age when the current spin period still remains close to the initial period and because a certain fraction of the pulsars are known to have unexpectedly slow spin periods which indicate preceding rapid decelerations of their spins (e.g., \citealp{nakano15, enoto17, enoto19}).
In the latter case, these discrepancies can be attributed to the decay of the magnetic field of the pulsars \citep{colpi00, dallosso12}.

We evaluate the possible systematic uncertainties of $t_{\rm kin, NS}$, considering the following three potential error sources: (1) selection bias in the SNR shapes (see Section~\ref{sec-kin}), (2) something in the data reduction procedure in Section~\ref{sec-kin}, and (3) possible offset between the geometric center and actual explosion point.

To investigate (1), we prepare the data with increased missing parts of the shell where the boundary between the SNR and ISM is relatively ambiguous, fit the data, and see the difference of the estimated position of the explosion center and resultant $t_{\rm kin, NS}$ from those obtained in Section~\ref{sec-kin}. Although the estimated explosion centers' errors are larger than those obtained in Section~\ref{sec-kin}, the resultant $t_{\rm kin, NS}$ is found not to differ significantly from the values presented in Table~\ref{tab-ages} presumably because the error of $t_{\rm kin,NS}$ is dominated by the uncertainty of the proper motion. 

To investigate (2), firstly, we modify the treatment of the statistical weight of the estimated geometric center coordinates in each level of the data reduction when filling the histograms of the explosion-point estimates (e.g., in the reduction level 3, each of the three estimates of the explosion-point coordinates had a statistical weight of 1/3, whereas it was unity in the analysis presented in Section~\ref{sec-kin}).
Secondly, we perform the data reduction of the SNR shapes on the basis of $\theta$ instead of the ID of the data points (e.g., in the reduction level 3, the data points used to estimate the position of the explosion satisfy $\theta \,({\rm degree}) =$ 0--1, 3--4, 6--7, ..., 357--358). Then we calculate the explosion center positions and $t_{\rm kin, NS}$, using this reduced shape data in the same way as described in Section~\ref{sec-kin}.
The resultant $t_{\rm kin, NS}$ again does not differ significantly from the values presented in Table~\ref{tab-ages} presumably for the same reason as in case (1).

As for (3), the effect of the possible offsets between the geometric centers and actual explosion points is considered.
In this case, $t_{\rm kin,NS}$ can be constrained only for CTB~80, IC~443, Monogem, S~147, W~41, and W~44, which have larger separation angles between the estimated explosion points and positions of the NSs than the possible offsets (Fig.~\ref{fig-snr-psr-fit}).
However, we find that $t_{\rm kin,NS}$ of these objects can still be constrained well and the errors of $t_{\rm kin,NS}$ are still smaller than a factor of four in most cases, so that the accuracy of a factor of four for $t_{\rm dyn}$ and $t_{\rm p}$ is not affected significantly.

In consequence, our result that $t_{\rm dyn}$ and $t_{\rm p}$ agree well with $t_{\rm r}$ within a factor of four is not affected significantly by these potential systematic uncertainties.

\section{Initial Spin Periods of Pulsars}\label{sec-period}
In Section~\ref{sec-comparison}, large deviations of $\tau_{\rm c}$ from $t_{\rm r}$ have been found in five cases (Fig.~\ref{fig-ages2}).
Under an assumption that these deviations primarily originate in $P(0)$ and that none of the pulsars experiences magnetic field decay, we calculate $P(0)$ of the pulsars in our sample from $t_{\rm r}$, using Eq.~\ref{eq-initialspin}.
Here we assume the simplest case $n=3$, as in \cite{popov12}, except for Crab and Vela, for which $n$ is known \citep{espinoza17} and is adopted in our calculations.

Table~\ref{tab-spin} lists the derived $P(0)$ for the pulsars. We successfully determine $P(0)$ for all but G292.0+1.8 and Monogem, of which the $t_{\rm r}$ turns out to be larger than $\tau_{\rm c}$. 
The parameter $P(0)$ is found to vary significantly among the objects.
They are mostly consistent with those presented in \cite{popov12}, but are more strictly constrained.
This is expected, given that they used the historical ages or $t_{\rm dyn}$ as the reliable ages, and $t_{\rm dyn}$ has been found to be reliable within a factor of four in Section~\ref{sec-comparison} in this work.

\begin{table*}[htb]
\centering
\caption{
Spin properties of the pulsars associated with the SNRs in our sample.
\label{tab-spin}}
\begin{threeparttable}
\begin{tabular}{l l r c l l}
\hline\hline
SNR & Pulsar & $P$\tnote{a} & $\dot{P}$\tnote{b} & $n$\tnote{c} & $P (0)$\tnote{d} \\
\hline
3C58 & J0205+6449 & 65.716 & $1.938 \times 10^{-13}$ & --- & 60.379 \\
Crab & B0531+21 & 33.392 & $4.210 \times 10^{-13}$ & 2.342$\pm$0.001 & (19$\pm 3) \times 10^{-3}$ \\
G11.2$-$0.3 & J1813$-$1749 & 44.699 & $1.266 \times 10^{-13}$ & --- & 37.626 \\
G292.0+1.8 & J1124$-$5916 & 135.477 & $7.526 \times 10^{-13}$ & --- & ---\tnote{e} \\
Puppis~A & J0821$-$4300 & 112.799 & $9.300 \times 10^{-18}$ & --- & (113$\pm 2) \times 10^{-4}$ \\
W44 & B1853+01 & 267.440 & $2.084 \times 10^{-13}$ & --- & 78$\pm$84 \\
Vela & B0833$-$45 & 89.328 & $1.250 \times 10^{-13}$ & 1.7$\pm$0.2 & 25$\pm$32 \\
G114.3+0.3 & B2334+61 & 495.370 & $1.934 \times 10^{-13}$ & --- & 358$\pm$24 \\
S147 & J0538+2817 & 143.158 & $3.669 \times 10^{-15}$ & --- & 129.3$\pm$4.2 \\
W41 & B1830$-$08 & 85.288 & $9.177 \times 10^{-15}$ & --- & 65.8$\pm$4.0 \\
CTB80 & B1951+32 & 39.531 & $5.845 \times 10^{-15}$ & --- & (29.4$\pm 6.6) \times 10^{-1}$ \\
Monogem & B0656+14 & 384.929 & $5.494 \times 10^{-14}$ & --- & ---\tnote{e} \\
\hline
\end{tabular}

\begin{tablenotes}
\item[a] Spin period of the pulsar (ms).
\item[b] Period derivative of the pulsar (s s$^{-1}$).
\item[c] Braking index taken from \cite{espinoza17}. If unknown, $n=3$ is assumed in the calculation of $P(0)$.
\item[d] Initial period of the pulsar (ms) derived in this work.
\item[e] $P (0)$ is not determined because the assumed age turns out to be larger than the present characteristic age $P/2\dot{P}$.
\item[]{\bf Notes.} $P$ and $\dot{P}$ are taken from the ATNF pulsar catalog (\url{http://www.atnf.csiro.au/research/pulsar/psrcat/}). Their errors are not included.

\end{tablenotes}
\end{threeparttable}
\end{table*}

\section{Summary}
In this study, we compared the dynamical and plasma ages of SNRs and characteristic ages of the associated pulsars with reliable ages ($t_{\rm r}$), namely one or more of the historical age, light-echo age, { kinematic age of the ejecta knots} and kinematic age of the associated NS.
We estimated the NS kinematic ages for eight systems by quantitatively determining the geometric centers of the SNR shells.
Consequently, the dynamical and plasma ages of the SNRs were found to agree with $t_{\rm r}$ within a factor of four, whereas the discrepancies of the characteristic ages of the associated pulsars with $t_{\rm r}$ are sometimes larger.
In addition, we also estimated the initial spin periods of the associated pulsars from the estimated kinematic ages with more strict constraints than the previous work.

\acknowledgments
The authors appreciate helpful comments by S. Katsuda about geometric center estimation methods and the ejecta kinematic ages, T. Holland-Ashford about geometric center estimation methods, O. Salvatore about simulations of SNR shapes, H. Odaka, T. Kasuga, and T. Tamba about the analysis procedure in geometric center estimation, and K. Makishima about the pulsar ages.
HS is supported by JSPS Research Fellowship for Young Scientist (No. 19J11069).
This research was partially supported by JSPS KAKENHI Grant Nos. 19K03908 (AB) and 18H01246 (SS), the Grant-in-Aid for Scientific Research on Innovative Areas ``Toward new frontiers: Encounter and synergy of 
state-of-the-art astronomical detectors and exotic quantum beams'' (18H05459; AB), and a Shiseido Female Researcher Science Grant (AB).





\bibliography{age_estimation_20201221}
\bibliographystyle{aasjournal}



\end{document}